 \documentclass[10pt,journal]{ieeeconf}
\usepackage{amssymb}
\usepackage[cmex10]{amsmath}
\usepackage{graphicx}
\usepackage{cite}
\usepackage{verbatim}
\usepackage[labelfont=bf]{caption}
\usepackage{subcaption}
\usepackage{authblk}
\usepackage{ulem,color}

\def\rr{\mathbf{r}}
\def\R{\mathbb{R}}

\def\R{\mathbb{R}}
\def\>{\rightarrow}
\def\({\left(}
\def\){\right)}

\newcommand{\norm}[1]{\left\|\, #1 \,\right\|}

\IEEEoverridecommandlockouts

\begin{document}

\title{Pattern-Acquisition in Finite, Heterogenous, Delay-Coupled Swarms}
\author{Klimka Szwaykowska$^{1}$*}
\author{Christoffer Heckman$^{1}$* \thanks{* Co-first author}}
\author{Luis Mier-y-Teran-Romero$^{1,2}$}
\author{Ira B. Schwartz$^{1}$}
\affil{$^{1}$ U.S. Naval Research Laboratory, Code 6792, Plasma Physics Division, Nonlinear 
Dynamical Systems Section, Washington, DC 20375 \\
$^{2}$ Bloomberg School of Public Health, Johns Hopkins University, Baltimore, MD 21205} 
\date{\today}
\maketitle

\begin{abstract}

The self-organizing behavior of swarms of interacting particles or agents is a
topic of intense research in fields extending from biology to physics and
robotics. In this paper, we carry out a systematic study of how the stable
spatio-temporal patterns of a swarm depend on the number of agents, in the
presence of time-delayed interaction and agent heterogeneity. We show
how a coherent pattern is modified as the number of agents
varies, including the time required to converge to it starting from random
initial conditions. We discuss the implications of our results for curve-tracking using autonomous robotic systems.

\end{abstract}


\section{Introduction}


The dynamics of interacting multi-agent or swarming systems in  many
biological, physical and engineering fields is undergoing very active
research. Remarkably, these systems are able to self-organize into highly
structured spatio-temporal patterns and demonstrate complex collective behaviors, with very limited 
information passed between individual agents. Examples in biology include bacterial colonies, 
schooling fish, flocking birds, swarming locusts, ants, and even pedestrians \cite{Budrene1995, 
Toner95, Parrish99, Topaz2004, Helbing1995,  Farrell12, Polezhaev2006, Tunstrom2013,Lee2006, 
Mishra12, Xue12}. 

One of the practical goals in swarming research in robotics is to bring the advantages of aggregation in biological 
systems, such as scalability and robustness, into engineered systems. To achieve scalability, the 
individual agents that make up the aggregate should be simple and inexpensive, and easily added to 
the overall swarm. For robustness, it should be possible to have agents fail and be removed from the 
swarm, without significantly affecting the collective dynamics. In fact, aggregates of locally 
interacting agents have been proposed as a means to create scalable sensor arrays for surveillance 
and exploration \cite{Bhatta2005,Wu2011,Leonard02,Justh2004b, Morgan05, chuang2007,Lynch2008}; and 
for the formation of reconfigurable modal systems, in which a group of simple agents can be used to 
accomplish a task that would be impossible for any agent individually 
\cite{Lynch2008,Kar2008,Dorigo2013}. 

A thorough understanding of the dynamical properties of the swarm is necessary
for algorithm design and implementation. Many different approaches are
possible in order to elucidate this: a number of works treat the swarm 
at a the level of individual agents
\cite{Helbing1995,Lee2006,Vicsek2006,Tunstrom2013}; others have attacked the
problem via continuum models
\cite{Edelstein-Keshet1998,Topaz2004,Polezhaev2006}. A number of studies show
that even with simple interaction protocols, swarms of agents are able to
converge to organized, coherent behaviors. Interestingly, environmental
noise and processing time delays affecting the agent dynamics can lead to the formation
of new steady-state motions, bistability and hysterisis, or phase transitions between between co-existing steady states 
\cite{MieryTeran2012,MieryTeran2011,MieryTeran2014}. Noise is used to model the
effects of external, unforeseen disturbances as well as unmodeled inter-agent
interactions, including uncertain coupling and/or communication. On the other hand, time delays are essential for modeling finite
communication and processing speeds in many interacting 
biological systems, including population dynamics, blood cell production,  
and genetic networks \cite{Martin2001,Bernard2004,Monk2003}, or in mathematical models of robot 
networks with explicit communication and processing delays \cite{Forgoston2008}.

Biology has shown us that swarms exist in stable configurations   composed  of a great many single agents. Motivated by this observation, a number of existing works on the spatio-temporal patterns of swarm
dynamics present results that are valid in the so-called ``thermodynamic limit,'' where the number of agents is
assumed to be very large {\cite{Vicsek2006,Edelstein-Keshet1998,Topaz2004,MieryTeran2012,MieryTeran2011,MieryTeran2014,Leverentz2009,Burger2013}. This limiting situation is attractive on several grounds and is particularly amenable to mean-field approximations that allow one to make analytic predictions on the collective behavior of the
swarm. However, in most real-world situations, the size of the swarm is
limited by the cost of individual agents, bandwidth requirements, agent
malfunction, etc. It is therefore important to verify how well the analytical
predictions for collective motion made in the thermodynamic limit hold as
swarm size is reduced.

In addition, most existing work assumes that the members of the swarm are
identical. However, many practical applications involve
swarms that are composed of agents with differing dynamical properties from
the onset, or that become different over time due to malfunction or
aging. Swarm heterogeneities lead to interesting new collective dynamics such as spontaneous 
segregation of the various populations within the swarm; it also has the potential to erode swarm cohesion. In biology, for example, it has been shown that sorting behavior of different cell types during the development of an organism can be 
achieved simply by introducing heterogeneity in inter-cell adhesion properties 
\cite{Steinberg1963,Graner1993}. It has also been shown that increasing the neighbor-to-neighbor attraction between cells of a single type leads to segregation of types in aggregates of self-propelled cells \cite{Belmonte2008}. In robotic systems, heterogeneity may arise over
time when, for example, battery  depletion, or other losses of functionality, occur at different rates for different agents within the swarm. Allowing for
heterogeneity in dynamical behaviors of swarm agents gives greater flexibility in system design, and is therefore desirable not only from a theoretical but also from a practical point of view.

In this work we address the issues of coherent-pattern scalability and robustness for a delay-coupled swarm with heterogeneous agents performing a path-following mission. We extend a globally delay-coupled swarm model in \cite{MieryTeran2011,MieryTeran2012} to swarms with heterogeneous agent dynamics. We conduct a careful numerical analysis to examine the scaling behavior of the coherent patterns of the swarm as the number of agents varies over a wide range. In particular, we
investigate how long it takes the swarm in an arbitrary configuration to
acquire a particular coherent pattern. This state exploits the segregation of heterogeneous agents in the swarm to create a state that separates agents according to their natural motion. The approach is novel in that it promote robustness by eliminating weaker agents that might negatively affect the performance of the swarm and serve as a metric for the health of the entire swarm.


\section{Problem Statement}
\label{sec:problemstatement}


{We investigate the pattern-transition capabilities of a two-dimensional swarm of
  autonomous agents as it carries out a
  mission. For definitive purposes, we take this mission to be the tracking of
  a virtual ``leader agent'' that moves along an arbitrary curve
  $\mathcal{C}$. We assume that each agent can measure the distance and
  relative heading to all other agents in the swarm and of the virtual
  leader. These measurements are relatively easy to obtain using inexpensive
  sensors, and do not require the agents to agree on a common reference frame.}

{At some point during the curve-tracking, the swarm is made to transition to a
  stationary, `ring' coherent pattern that may serve as a diagnostic state to identify
  the agents with degraded performance. Of particular interest is how the
  spatio-temporal scales of this diagnostic state depend on the number of
  agents present as well as the time to acquire said pattern.}

We now introduce the dynamical model for the swarming agents and the virtual
leader. Let $\mathbf{r}_i(t) \in \R^2$ denote the position of agent $i$ in the
swarm, $i\in \{1,\ldots,N\}$. Each agent has self-propulsion with a preferred
speed that is scaled to $1$, and additionally is attracted to the other agents
in the swarm and to the leader. The coupling coefficient that quantifies the
attraction between agents in the swarm is $a$, while the coupling coefficient
between each swarm agent and the leader is $a_L$. We consider that the
attraction between agents occurs in a time-delayed fashion, on account of
finite communication speeds and processing times; we assume a single, fixed time delay denoted by
$\tau$. The position of the virtual leader is given by $\rr_L(t)$. The leader
is confined to stay on the curve $\mathcal{C}$. Let $s \in \R$ denote the speed of the leader
along $\mathcal{C}$. The dynamics of the swarm particles and leader are
described by the following dimensionless governing equations: 
\begin{align}
\ddot{\mathbf{r}}_i &= \kappa_i\left(1 - \norm{\dot{\mathbf{r}}_i}^2\right)\dot{\mathbf{r}}_i -
\frac{a \kappa_i }{N}\mathop{\sum_{j=1}^N}_{i\neq j}(\mathbf{r}_i(t) -
\mathbf{r}_j(t-\tau)) \notag \\
& \quad + a_L \kappa_i (\rr_i(t)-\rr_L(t)) \label{eq:drdt} \\
\dot{s} &= (1-s) - a_0\norm{\mathbf{r}_L(t) - \mathbf{R}(t)}, \label{eq:dsdt}
\end{align}
where dots are used to denote differentiation with respect to time. The self-propulsion of agent $i$ 
is modeled by the term
$\left(1 - \norm{\dot{\mathbf{r}}_i}^2\right)\dot{\mathbf{r}}_i$. At time $t$,
agent $i$ is attracted to the position of agent $j$ at the past time $t-\tau$
and to the current position of the leader. The factor $\kappa_i \in (0,1]$ scales
the acceleration of agent $i$, behaving like an inverse mass. One may interpret it
as a measure of the battery state of agent $i$ or some other source of
heterogeneity that impedes acceleration. We assume that for a given
number of agents in the swarm, $\kappa_i$ is given by a uniform distribution
on $(0,1]$. This simplified model does not 
include short-range repulsion or other collision-avoidance strategies; however, earlier studies 
with homogeneous swarms indicate that the collective dynamics of the swarm are not significantly 
altered by the introduction of short-range repulsion terms \cite{MieryTeran2012}.

We will show, using simulation, that the controller described above tracks the position of the 
virtual leader agent along the curve. We will further show that, to achieve segregation of the 
swarming agents by $\kappa$, it is sufficient to set the speed of the leader agent to $0$, for 
appropriate values of the parameters $a$ and $\tau$. For application of our theoretical results in a 
real-world setting (where number of agents is typically limited by space, cost, or communication 
bandwidth requirements), we run extensive numerical simulations to examine how collective behavior 
of the swarm depends on the number of its constituent agents $N$.


\section{Curve Tracking}
\label{sec:CT}


The goal of the curve-tracking behavior is to have the center of mass of the swarm track the 
length-parametrized curve $\mathcal{C}$. To this end, we introduce the virtual leader agent, with 
position $\rr_L(t) \in \mathcal{C}$, and add a proportional control term $a_L \kappa_i 
(\rr_i(t)-\rr_L(t))$ in (\ref{eq:drdt}) to track the position of this agent. 

The acceleration of the virtual leader along $\mathcal{C}$ is given by (\ref{eq:dsdt}), and 
includes a self-propulsion term as well as a feedback term that reduces acceleration proportionally 
to the distance from the virtual leader position the the swarm center of mass. A series of 
snapshots of the swarm tracking the virtual leader along a circular trajectory are shown in Fig. 
\ref{fig:snapshots}.

\begin{figure}[htb]
 \centering
 \begin{subfigure}{0.26\textwidth}
\includegraphics[width=\textwidth]{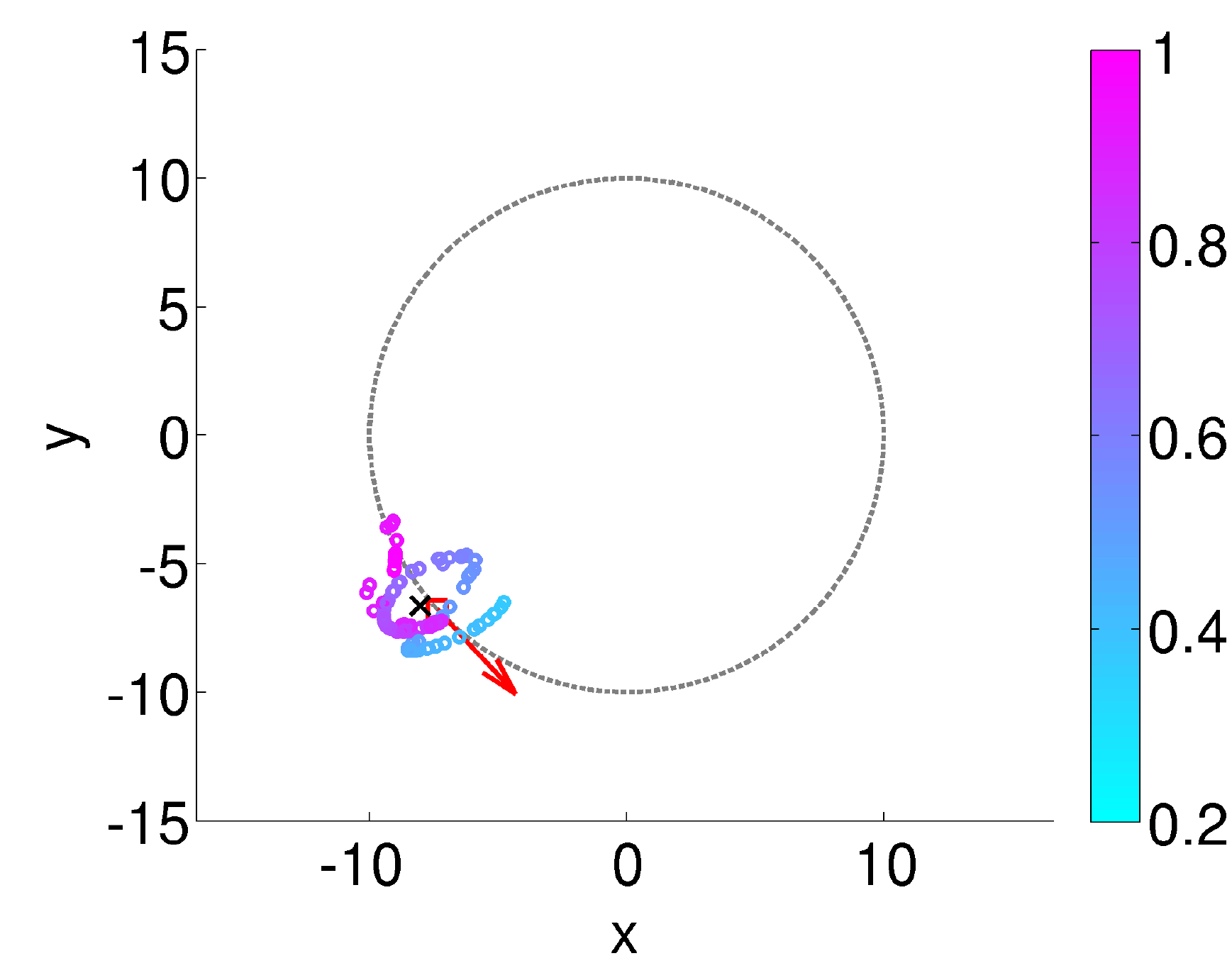} 
\vspace{-15pt}
\caption{$t = 81.8$}
\vspace{10pt}
\end{subfigure}
\begin{subfigure}{0.26\textwidth} 
\includegraphics[width=\textwidth]{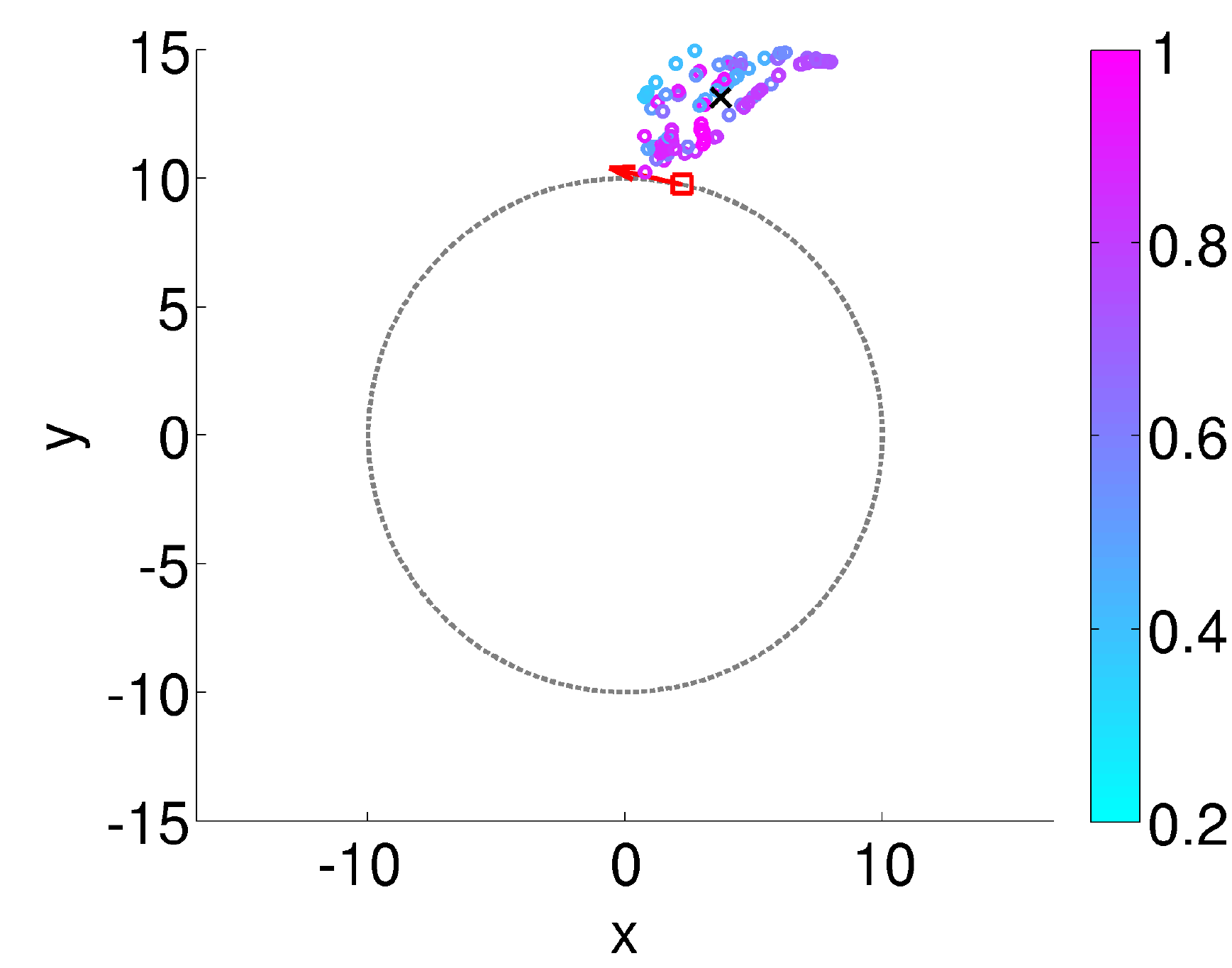}
\vspace{-15pt}
\caption{$t = 144.8$}
\vspace{10pt}
\end{subfigure}
\begin{subfigure}{0.26\textwidth}
\includegraphics[width=\textwidth]{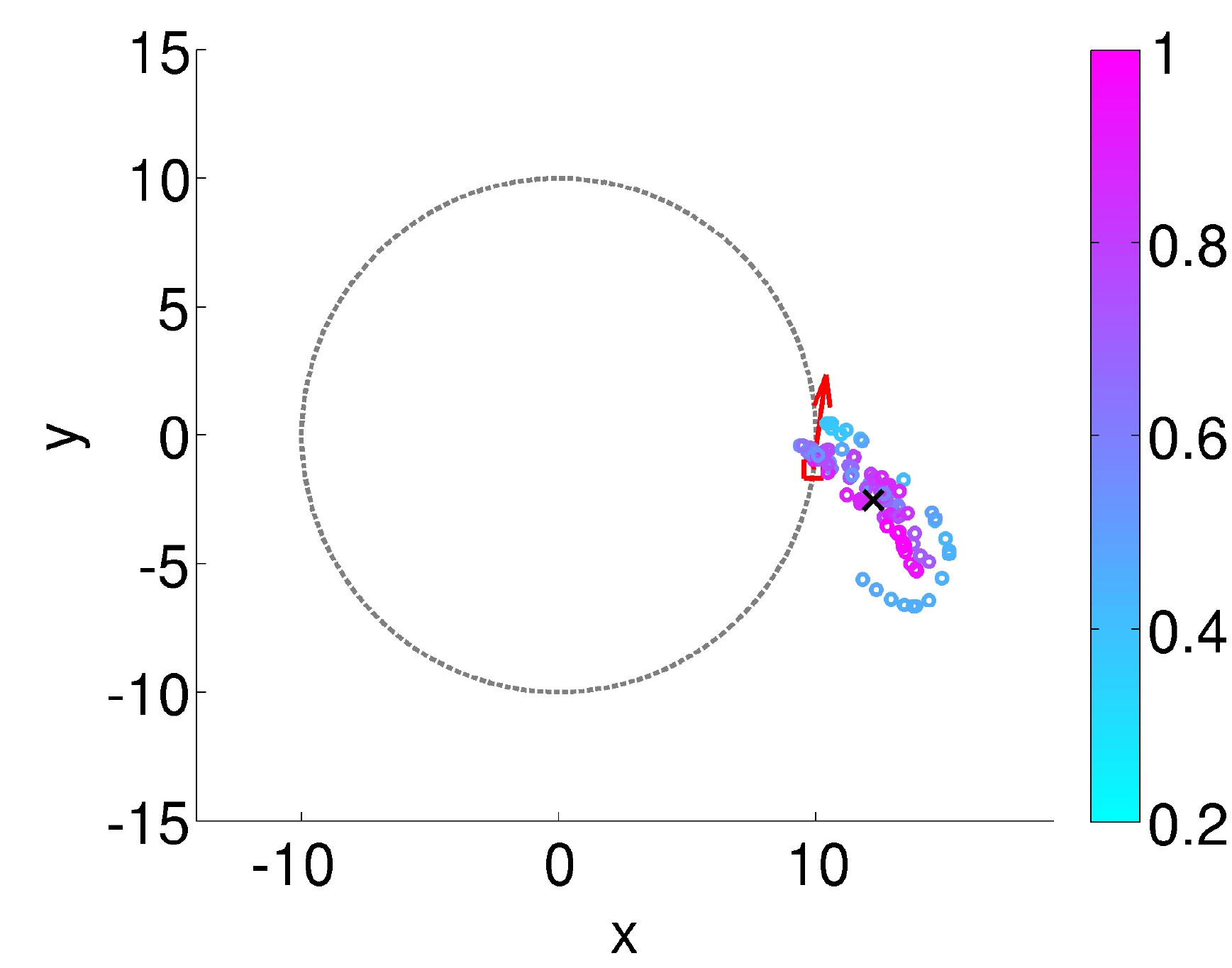} 
\vspace{-15pt}
\caption{$t = 228.8$}
\vspace{10pt}
\end{subfigure}
\caption{Snapshots from simulation of swarm tracking a circular trajectory. The position of the 
virtual leader is shown by the red box; the red arrow indicates the direction of the leader 
agent's motion. The center of mass of the swarm is marked by a black $\boldsymbol{\times}$. Agent positions are shown by circles, with colors indicating the respective values of $\kappa_i$. 
The values of $\kappa_i$ in this simulation are uniformly distributed on $[0.2,1]$. The coupling 
constant $a=0.1$, $a_L = 0.06$, and the delay is $\tau = 1$. The coupling coefficient of the leader 
to the swarm is $a_0 = 0.06$.}
\label{fig:snapshots}
\end{figure}


\section{Diagnostic Mode}


The system is switched into diagnostic mode by setting the leader speed $s=0$. When this happens, the swarm converges to a ring state, in which the center of mass is fixed at the position of the leader, while particles in the swarm rotate in either direction about the center, with radius that depends on the coupling constants $a$ and $a_L$, and on the individual constant of acceleration $\kappa_i$.

In this section, we conduct a careful theoretical analysis of the center of mass dynamics of the swarm in the thermodynamic limit ($N \> \infty$), extending results presented for the leaderless, homogeneous swarm case ($a_L=0$ and $\kappa_i = 1, \: \forall i \in \{1,\ldots,N\}$) in \cite{MieryTeran2012}.

Proceeding as in \cite{MieryTeran2012}, let $\mathbf{R}(t) = \frac{1}{N}\sum_{i=1}^N \rr_i(t)$ be 
the center of mass of the swarm, and let $\delta \rr_i(t) = \rr_i(t) - \mathbf{R}(t)$ be the 
position of agent $i$ relative to the center of mass; note that $\sum_{i=1}^N \delta \rr_i(t) = 0$. 
Using the change of variables $\rr_i(t) = \mathbf{R}(t) + \delta \rr_i(t)$, (\ref{eq:drdt}) can be 
written as
\begin{align}
\ddot{\mathbf{R}} + \delta\ddot{\rr}_i \!&=\! \kappa_i\left(\!\! 1 - \norm{\dot{\mathbf{R}}}^2 - \norm{\delta \dot{\rr}_i}^2 - 2 \langle \dot{\mathbf{R}}, \delta \dot{\rr}_i \rangle \!\!\right)\!\!(\dot{\mathbf{R}} + \delta \dot{\rr}_i) \notag \\
& -\!\! \frac{a \kappa_i}{N} \! \sum_{j \neq i}\!\left( \mathbf{R}(t) + \delta\rr_i(t) - \mathbf{R}(t-\tau)- \delta\rr_j(t-\tau) \right) \notag \\
& - a_L \kappa_i(\mathbf{R}(t) + \delta\rr_i(t) - \mathbf{R}(t) - \delta\rr_L(t)). \label{eq:dRpdelrdt}
\end{align}
Without loss of generality, we set the stationary leader position $\rr_L$ to 0. Following the approach in \cite{MieryTeran2012}, we sum (\ref{eq:dRpdelrdt}) over $i$, then take the limit as $N \> \infty$ and neglect all terms in $\delta \rr$ (see \cite{MieryTeran2012} for a justification of this simplification). The resulting equation for the motion of the center of mass of the swarm in the thermodynamic limit is:
\begin{equation}
\ddot{\mathbf{R}} = \bar{\kappa} \left( 1-\norm{\dot{\mathbf{R}}}^2 \right) \dot{\mathbf{R}} - (a + 
a_L)\bar{\kappa} \mathbf{R}(t) + a \bar{\kappa} \mathbf{R}(t-\tau),
\end{equation}
where $\bar{\kappa} = \frac{1}{N} \sum_{i=1}^N \kappa_i$ is the mean acceleration factor. The above system has an equilibrium point 
at $\mathbf{R} = \dot{\mathbf{R}} = 0$, which corresponds to the ring state. As $a$ and $\tau$ 
increase 
(for fixed values of the parameters $a_L$ and $\bar{\kappa}$), the system undergoes a Hopf 
bifurcation, 
giving rise to a new oscillating steady-state behavior, whereby the swarm
  becomes more compact ad organizes itself in to a coherent rotating state.  To ensure that the system converges to the 
ring state given $\tau$ and estimated $\bar{\kappa}$, we choose $a$ and $a_L$ so that the system 
lies below 
the first Hopf bifurcation curve (see Fig. \ref{fig:bifurcations}).

\begin{figure*}[htb]
 \centering
\begin{subfigure}[b]{0.35\textwidth} 
\includegraphics[width=\textwidth]{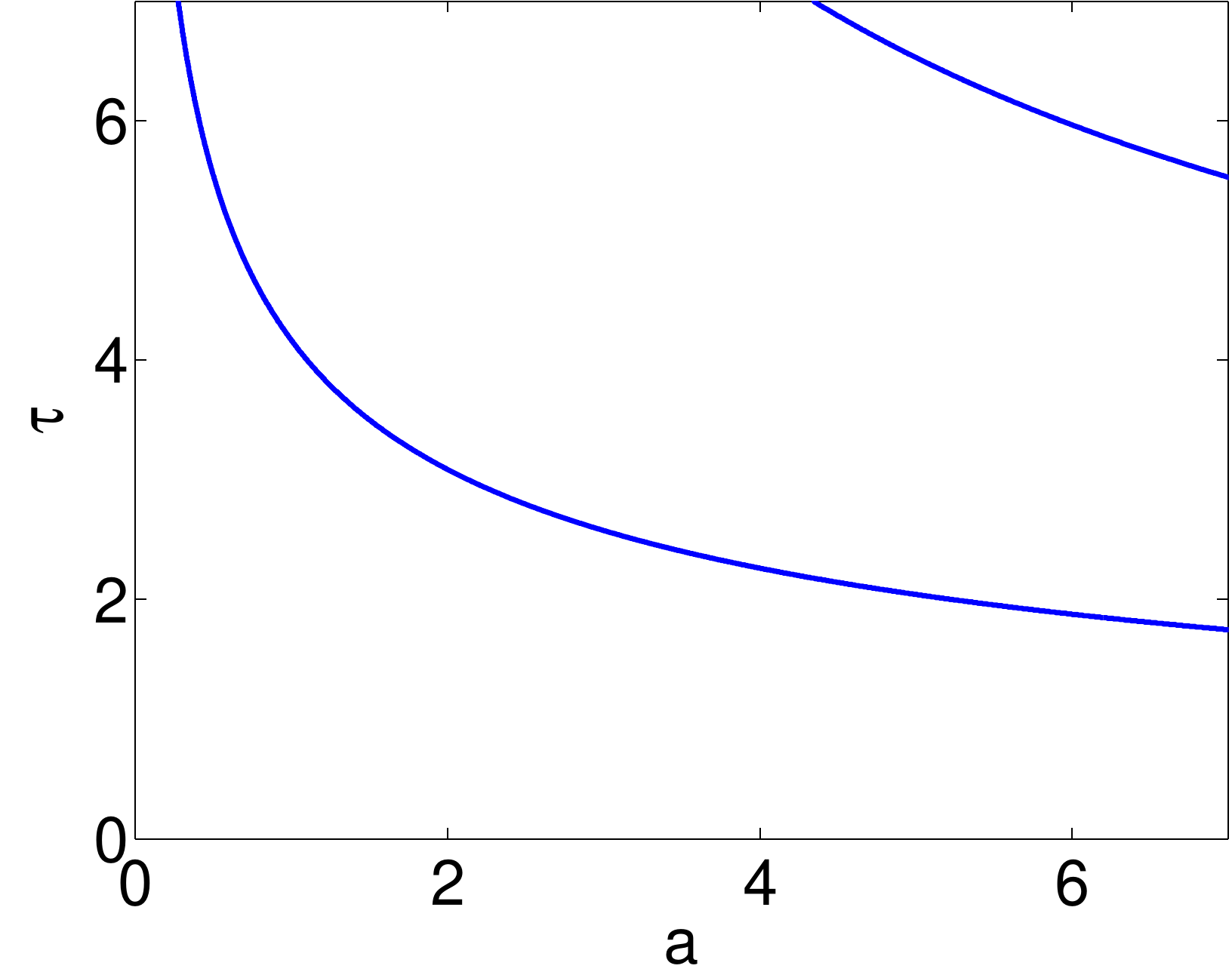}
\vspace{-15pt}
\caption{$\bar{\kappa} = 0.2$, $a_L = 0$ (leaderless case)}
\vspace{10pt}
\end{subfigure}
\begin{subfigure}[b]{0.35\textwidth}
\includegraphics[width=\textwidth]{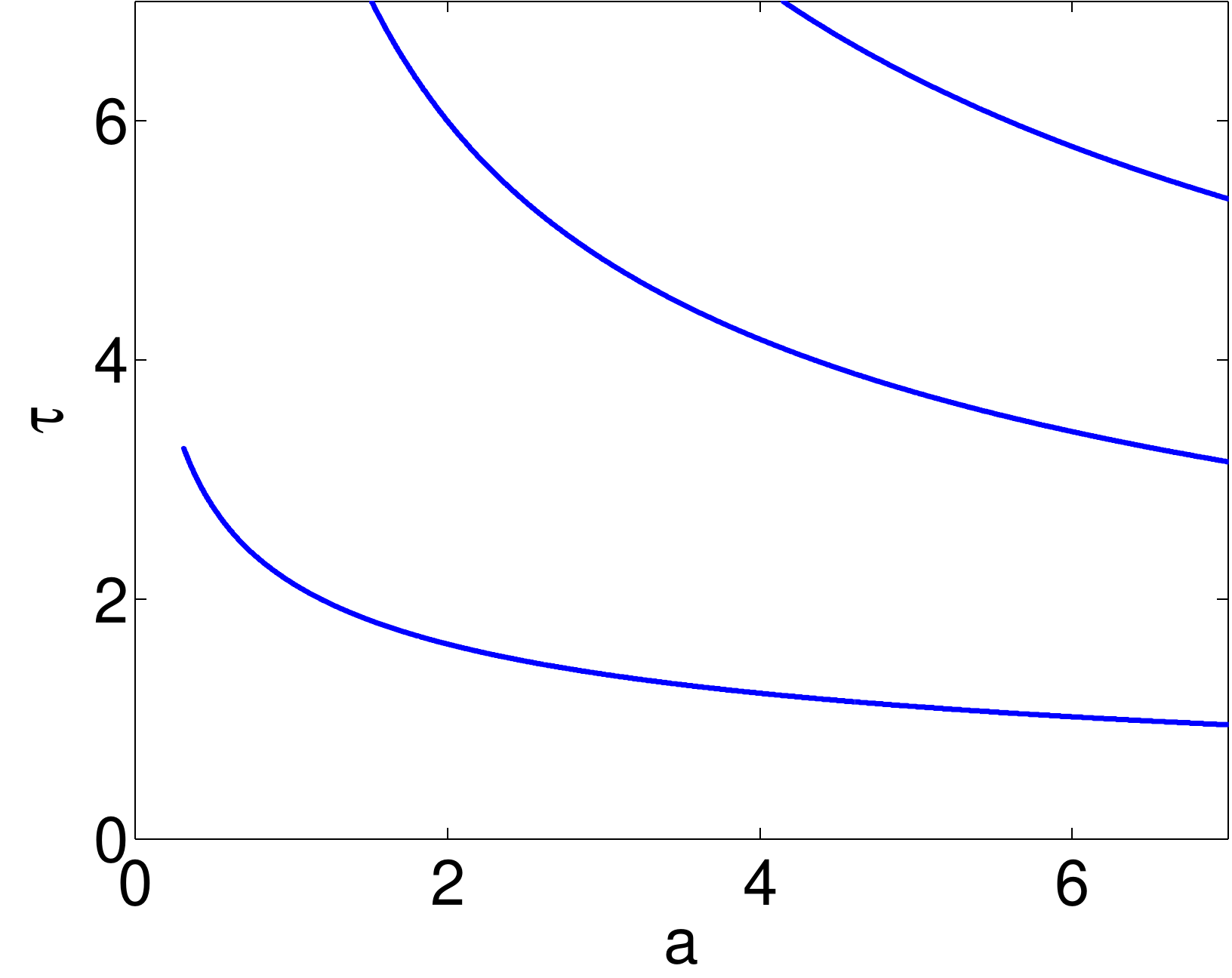} 
\vspace{-15pt}
\caption{$\bar{\kappa} = 0.61$, $a_L = 0$ (leaderless case)}
\vspace{10pt}
\end{subfigure}
\begin{subfigure}[b]{0.35\textwidth} 
\includegraphics[width=\textwidth]{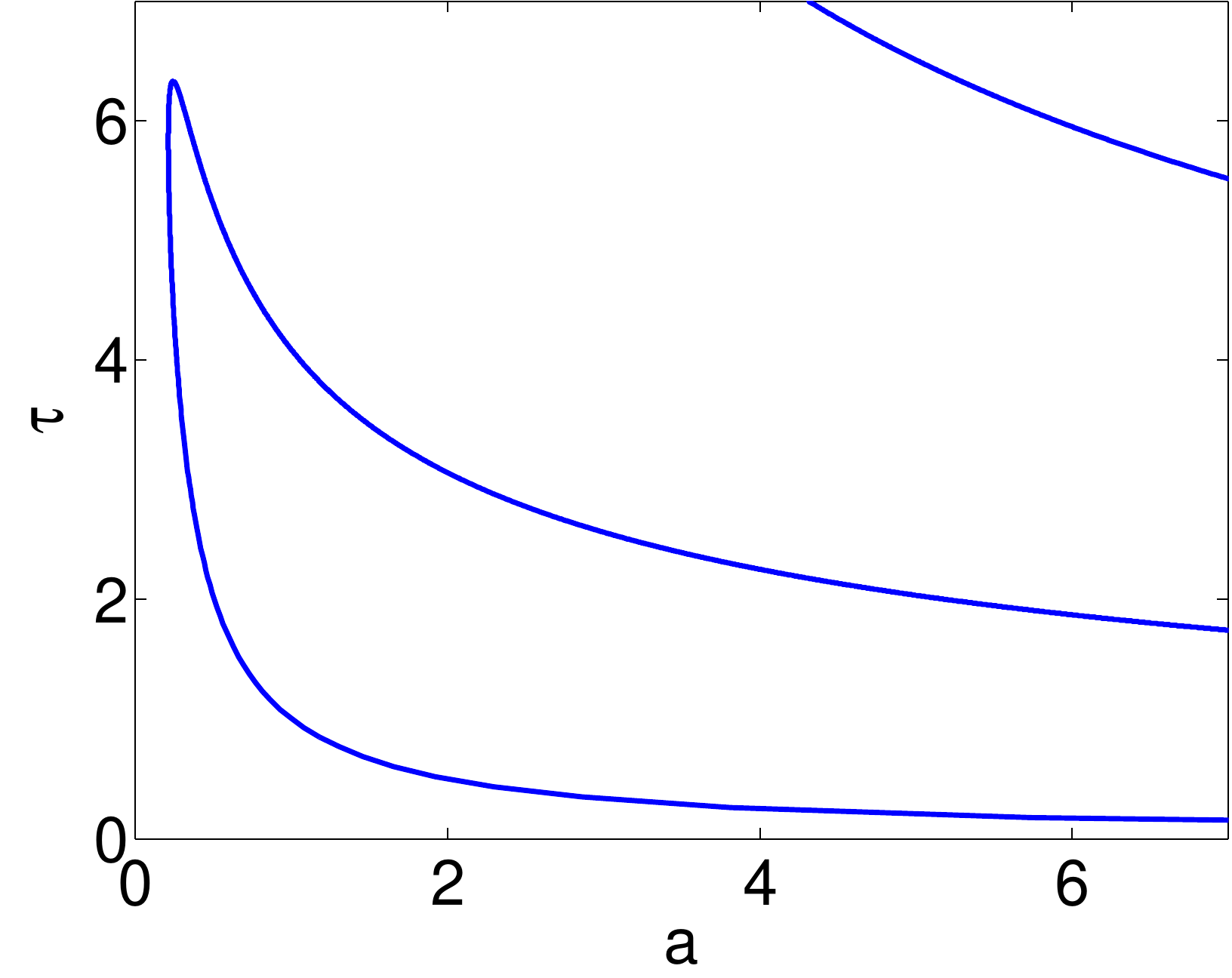}
\vspace{-15pt}
\caption{$\bar{\kappa} = 0.2$, $a_L = 0.06$}
\vspace{10pt}
\end{subfigure}
\begin{subfigure}[b]{0.35\textwidth}
\includegraphics[width=\textwidth]{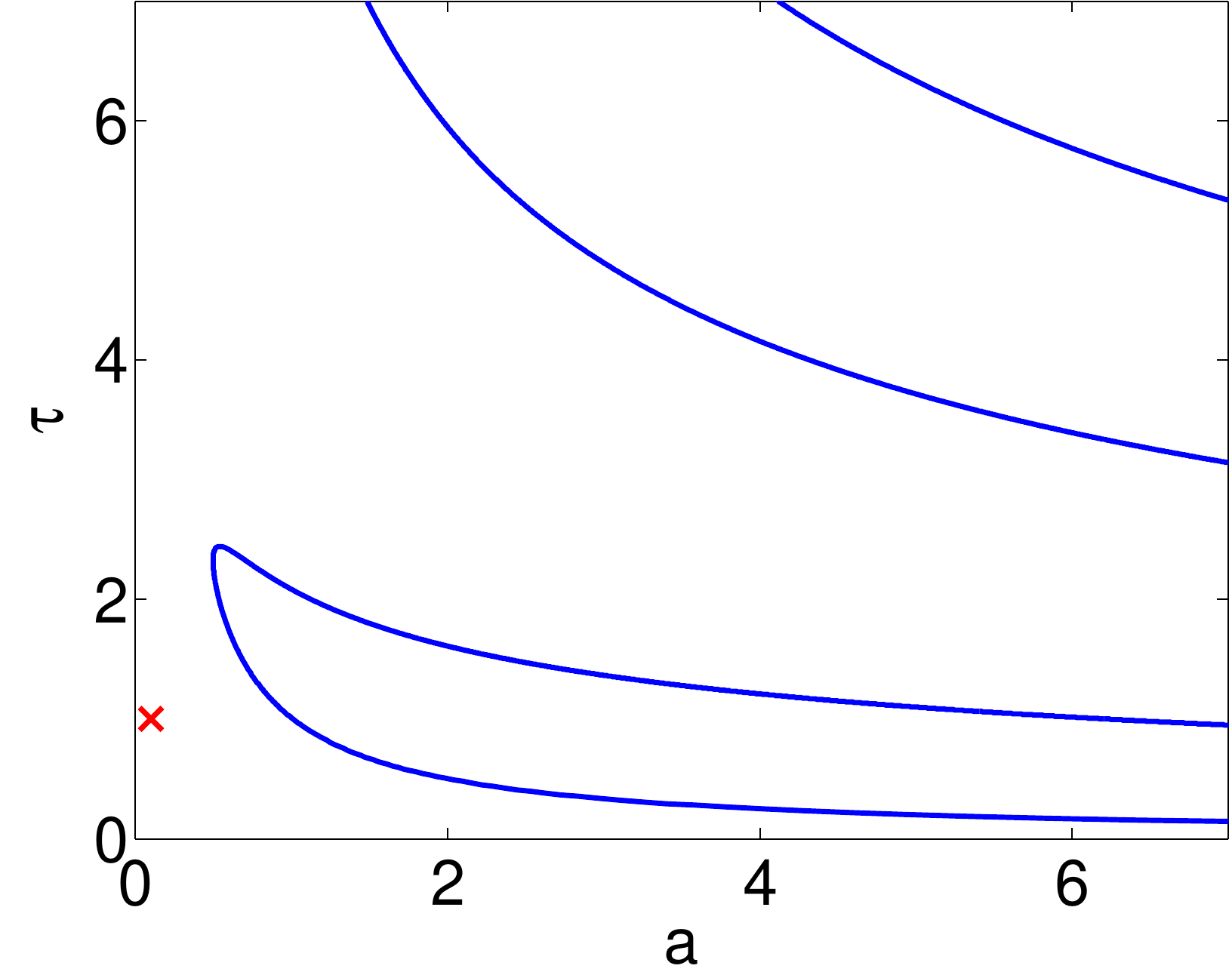} 
\vspace{-15pt}
\caption{$\bar{\kappa} = 0.61$, $a_L = 0.06$}
\label{subfig:simspace}
\vspace{10pt}
\end{subfigure}
\caption{The solid blue lines show $\tau$ vs $a$ Hopf bifurcation curves for the center-of-mass 
heterogeneous swarm dynamics, for different values of the parameters $\bar{\kappa}$ and $a_L$. The 
simulation in Section \ref{sec:CT} were run with parameter values marked by the red 
$\boldsymbol{\times}$ in Fig. \ref{subfig:simspace}.}
\label{fig:bifurcations}
\end{figure*}

Once the system enters the ring state, the agents with low battery power (denoted in our model by 
low $\kappa_i$) can be picked out from the radius of the circular trajectory they follow about the 
fixed center of mass. To see this, consider (\ref{eq:dRpdelrdt}). As before, we set $\rr_L = 0$, so that in the ring state, $\mathbf{R} = \dot{\mathbf{R}} = 0$. Equation 
(\ref{eq:dRpdelrdt}) then simplifies to:
\begin{align}
\delta\ddot{\rr}_i &= \kappa_i \left( 1-\norm{\delta \dot{\rr}_i}^2 \right)\delta \dot{\rr}_i \notag \\
&- \frac{a \kappa_i}{N} \sum_{j\neq i} (\delta \rr_i(t) - \delta \rr_j(t-\tau)) - a_L \kappa_i \delta \rr_i(t).
\end{align}
Taking the limit as $N \> \infty$ and converting to polar coordinates $(\rho_i,\theta_i)$, where 
$\rho_i$ denotes the distance of agent $i$ from the origin and $\theta_i$ denotes the phase, gives
\begin{align}
\ddot{\rho}_i &= \kappa_i (1-\rho_i^2\dot{\theta}_i^2 - \dot{\rho}_i^2) \dot{\rho}_i + (\dot{\theta}_i^2 - (a+a_L)\kappa_i)\rho_i \\
\rho_i \ddot{\theta}_i &= \kappa_i (1-\rho_i^2\dot{\theta}_i^2 - \dot{\rho}_i^2) \rho_i \dot{\theta}_i-2\dot{\rho}_i\dot{\theta}_i.
\end{align}
In the ring state, $\ddot{\rho}_i = \ddot{\theta}_i = \dot{\rho_i} = 0$; setting these values in the above equations gives 
\begin{align}
\dot{\theta}_i &= \pm \sqrt{(a+a_L)\kappa_i} \\
\rho_i &= 1/|\dot{\theta}_i| = 1/\sqrt{(a+a_L)\kappa_i}.
\label{eq:rhoi}
\end{align}

Thus, in the ring state, agent $i$ circles the stationary leader position with radius inversely proportional to $\kappa_i$. As a result, we can easily identify the agents with limited motion capabilities by distinguishing the agents whose radius in the ring state exceeds a certain pre-specified threshold. As an example, Figure \ref{fig:diagnostic} shows the case in which we wish to eliminate all agents with $\kappa_i 
\leq 0.5$; the threshold radius is then $\rho_{\rm TH} = 1/\sqrt{0.5(a+a_L)}$ as shown in the figure. 

\begin{figure}[htb]
 \centering
\includegraphics[width=0.4\textwidth]{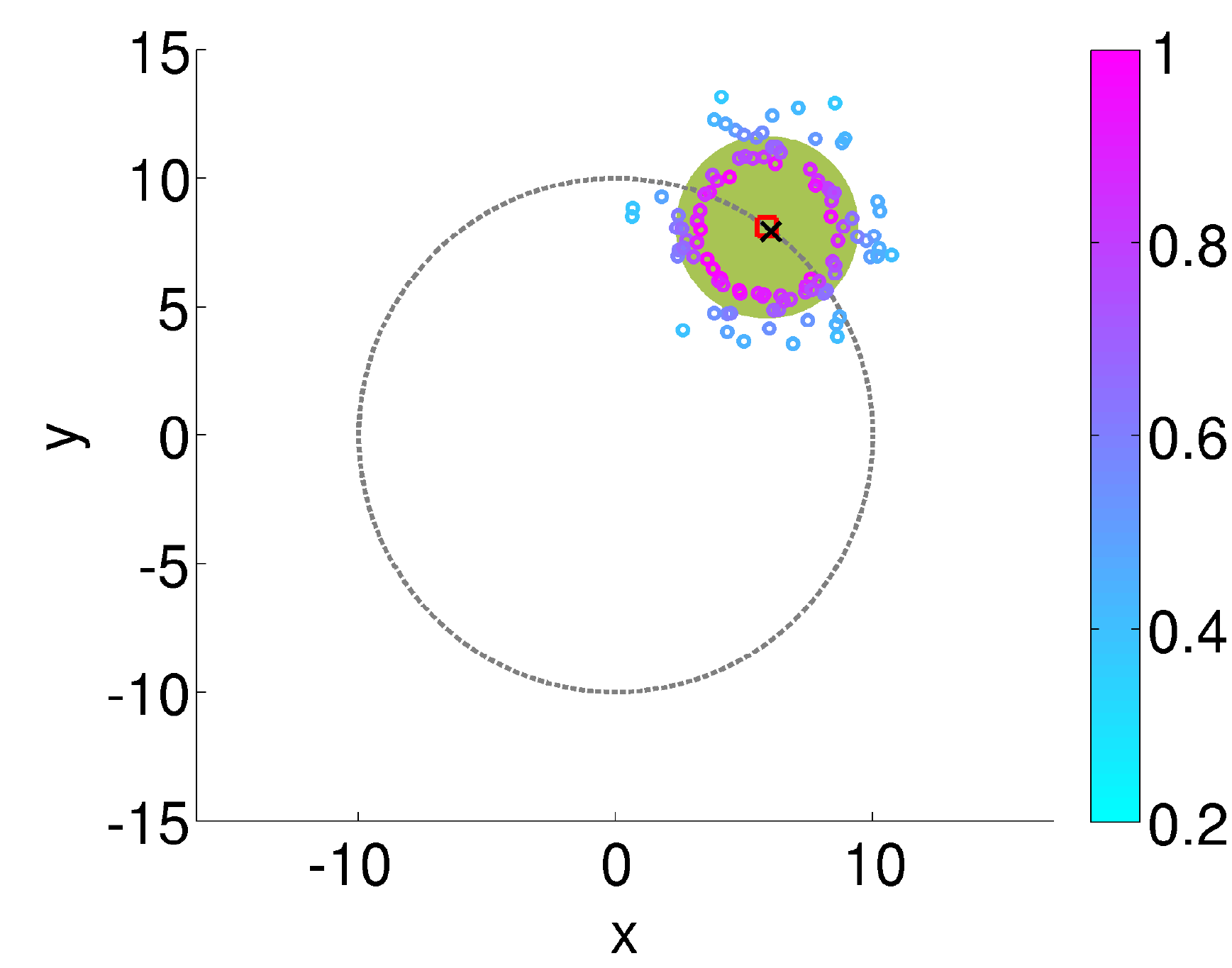} 
\caption{Snapshot from simulation of swarm in diagnostic state. The virtual leader is marked by the 
red box; the center of mass of the swarm is marked by a black $\boldsymbol{\times}$ and is 
coincident with the leader position. Agent positions are shown by circles, with colors 
indicating the respective values of $\kappa_i$. All agents with $\kappa_i > 
0.5$ are contained inside the green circle; agents which lie outside may be eliminated.}
\label{fig:diagnostic}
\end{figure}


\section{Finite $N$ Effects}

We now investigate how the spatio-temporal scales of the ring state, as well as how the length of time required to achieve it varies with the number of agents $N$. We consider the system to be in the ring state once the swarm's mean radius to its center of mass has converged to a small neighborhood about the average value.

Firstly, we set $\kappa_i=1$ for all agents and study the dependence of swarm
collective motions on finite $N$. Using numerical simulations, we have
measured the time required to converge to the ring state over 100 trials for
random initial conditions at various population sizes, ranging from $N=2$ to
$N=150$. As expected, for large $N$, the dynamics converge to a `thermodynamic limit' and are qualitatively similar in almost all trials. Fig. \ref{fig:timetoconvergea} shows that for large population
sizes the times to converge are relatively constant, but as $N$ decreases the
convergence time and the variance of these times increases dramatically. For very small $N$ the system is far less predictable and 
will at times converge to a periodic motion different from the ring state.

When agents do converge to the ring state, we can make the following theoretical prediction for the radius $\rho$ of the ring in the finite-$N$ case, under the assumption that agents are uniformly distributed along the ring: 

\begin{align}
\omega^2 &= a \left(1 - \frac{1}{2}(1-\cos \omega \tau) \right),\\
\rho &= \frac{1}{\omega} \sqrt{1 + \frac{a \sin \omega \tau}{N \omega}}.
\end{align}

\noindent where $\omega$ is the angular frequency of the agents moving about the ring. For $N \rightarrow \infty$ these reduce to $\omega^2 = a$ and $\rho =  \frac{1}{\omega}$,
which agrees with Eq.\ \eqref{eq:rhoi} for the ring radius in the
large $N$ limit, remembering that $a_L=0$ and $\kappa_i = 1$ in the current situation. In this case of homogeneous $\kappa$, Figure \ref{fig:homogeneous_swarm} shows good agreement 
between the simulated radius and velocity and the predictions of these
quantities as given above.

For the uniform $\kappa$ case, When $N$ is very small (less than 10), a wealth of behavior emerges that is far less likely for large
$N$. For example, when $N = 5$ the most prevalent state arranged all five agents equally along a
circle in a pentagonal pattern, rotating in the same direction. These symmetric patterns exhibit extremely
rapid convergence times from random initial conditions, which explains the surprising rapid decrease in convergence times for small $N$ shown in Fig. \ref{fig:timetoconvergea}. Formally identifying these
patterns and justifying their unique behavior is an area of future work that we plan to
investigate.

Next we consider the case of distributed $\kappa_i$ as described in 
Section \ref{sec:problemstatement}. We repeat the numerical simulations conducted for $\kappa_i=1$ to measure the time required to 
converge to an apparent ring state over 100 trials with randomly distributed initial conditions 
and $\kappa_i$ uniformly distributed over $[0.2,1]$ in each trial over various values of $N$. The results of these simulations are shown in Figure \ref{fig:timetoconvergeb}, demonstrating a similar relationship of 
time-to-convergence with $N$ as in the uniform $\kappa$ case.  The mean radius of the ring converges to approximately $1.4$, which is consistent with the mean-field prediction for $a=1$, $a_L=0$, and $\kappa_i$ uniformly distributed on $[0.2,1]$. Fig. \ref{fig:heterogeneous_swarm} shows the mean ring radius and velocity of agents in the ring state.


\begin{figure}
 \centering
\begin{subfigure}[b]{0.35\textwidth} 
\includegraphics[width=\textwidth]{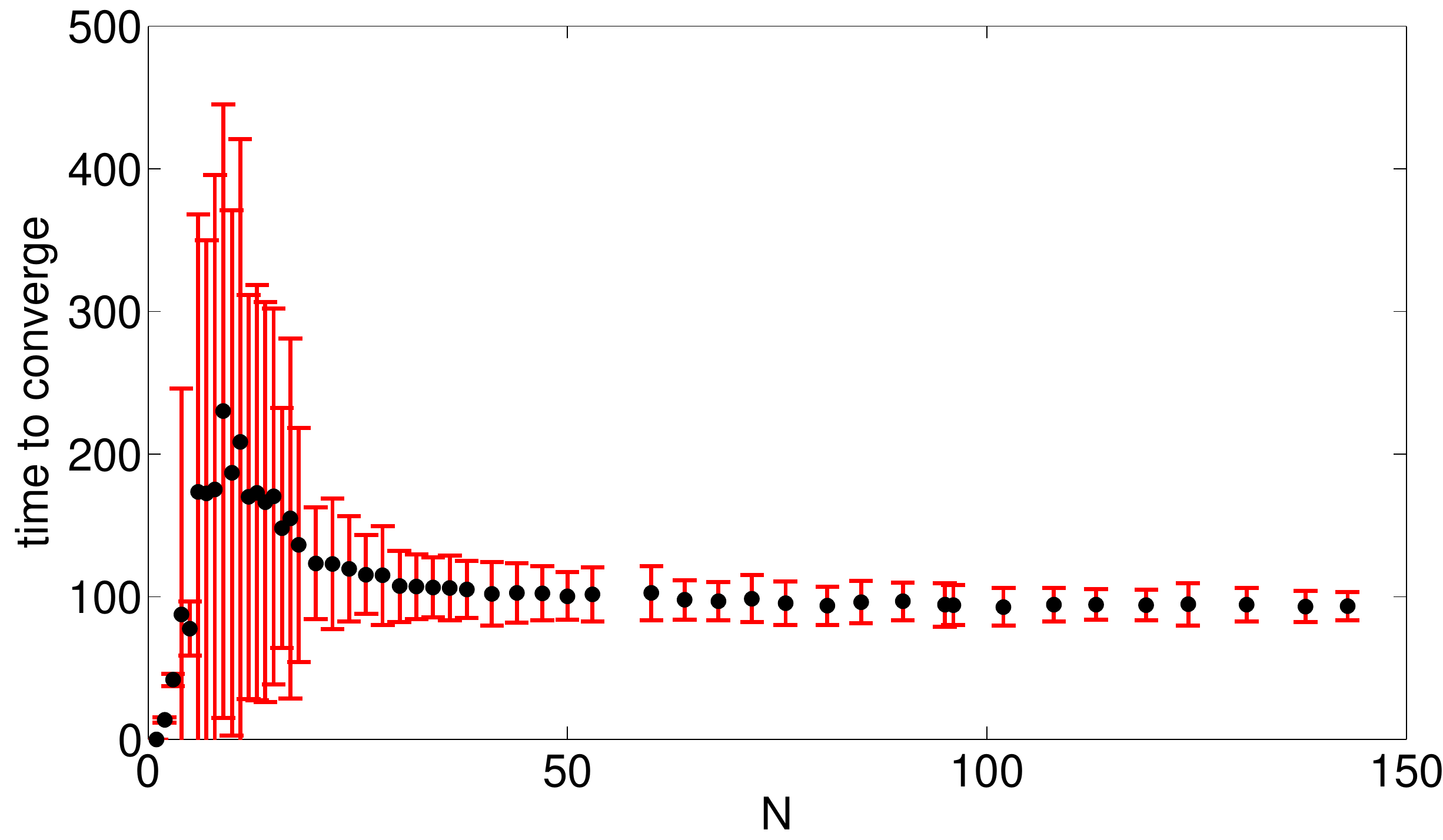}
\vspace{-20pt}
\caption{$\kappa_i = \bar{\kappa}$}
\label{fig:timetoconvergea}
\end{subfigure}
\begin{subfigure}[b]{0.35\textwidth} 
\includegraphics[width=\textwidth]{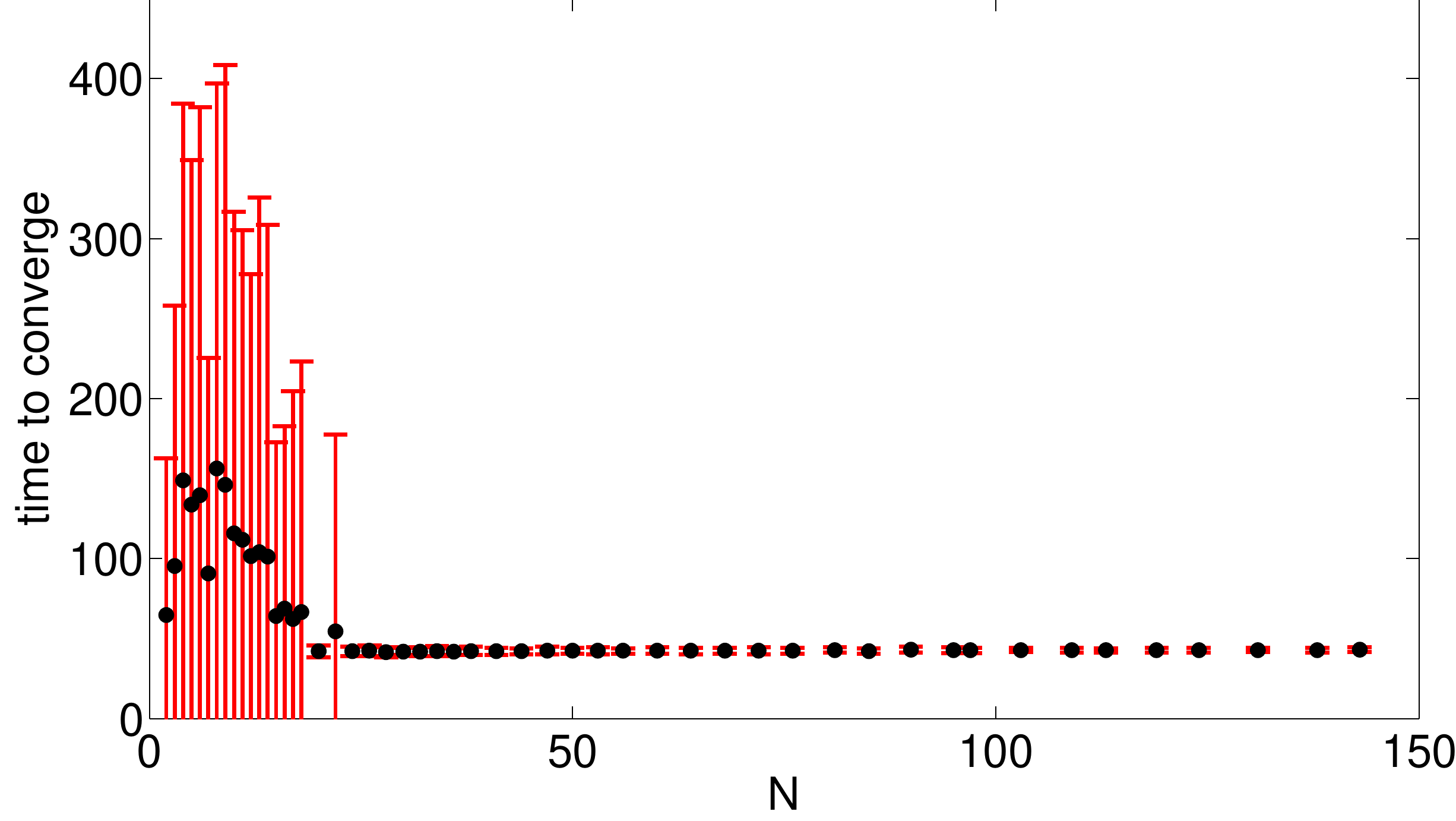}
\vspace{-20pt}
\caption{$\bar{\kappa} = 0.6$, uniform distribution} 
\label{fig:timetoconvergeb}
\vspace{10pt}
\end{subfigure}
\caption{Time for the system to converge to ring state for different values of swarm size $N$. 
Parameter values are $a = 1$, $a_L = 0$, $\tau = 1$ and  $\kappa_i = 1$ for all
$i$ (top panel) and 
$\kappa_i$ uniformly distributed on $[0.2,1]$ (bottom panel).}
\label{fig:timetoconverge}
\end{figure}

\begin{figure}[ht!]
 \centering
\begin{subfigure}{0.45\textwidth} 
\includegraphics[width=\textwidth]{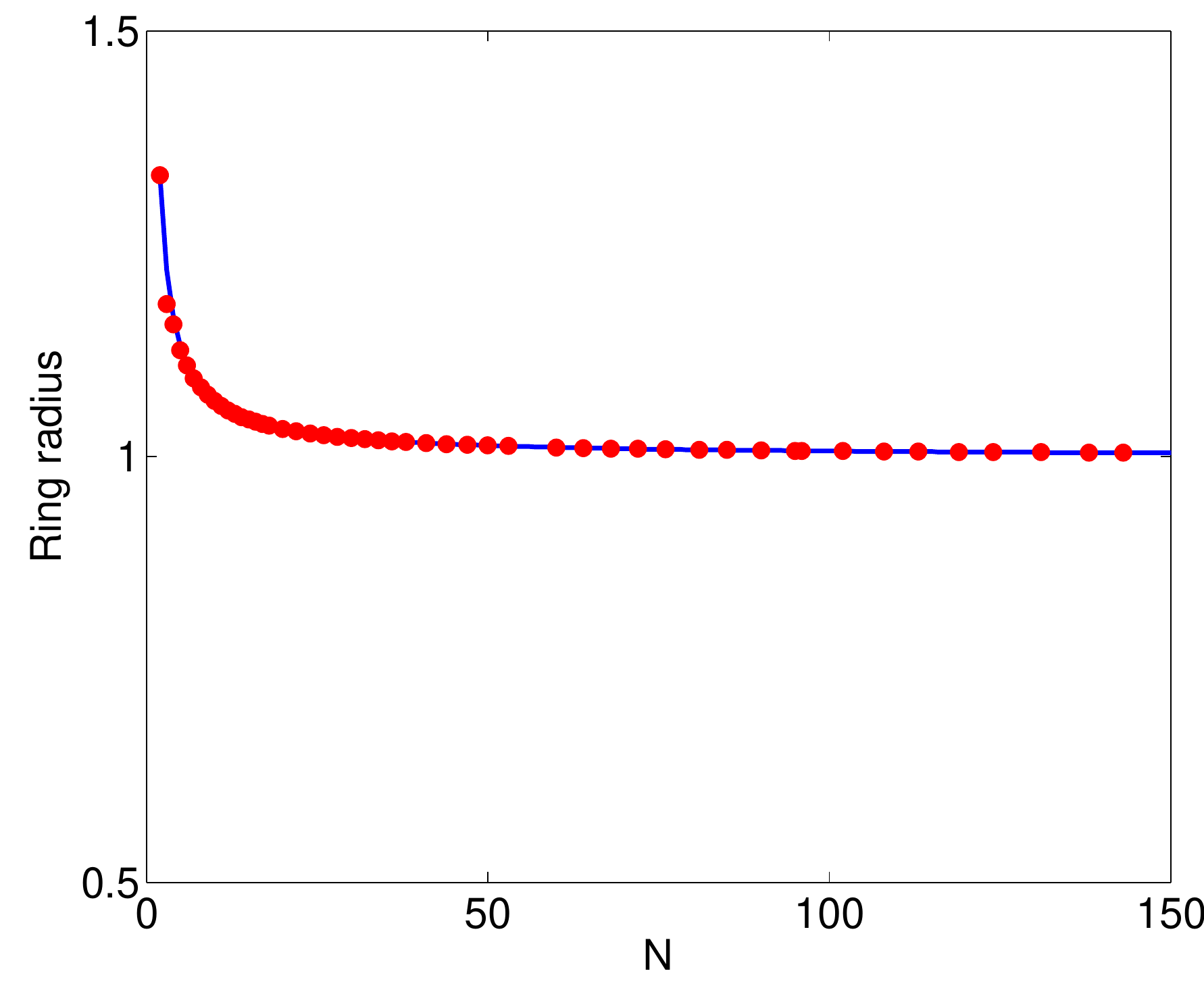}\\
\vspace{-10pt}
\end{subfigure}
\begin{subfigure}{0.45\textwidth} 
\includegraphics[width=\textwidth]{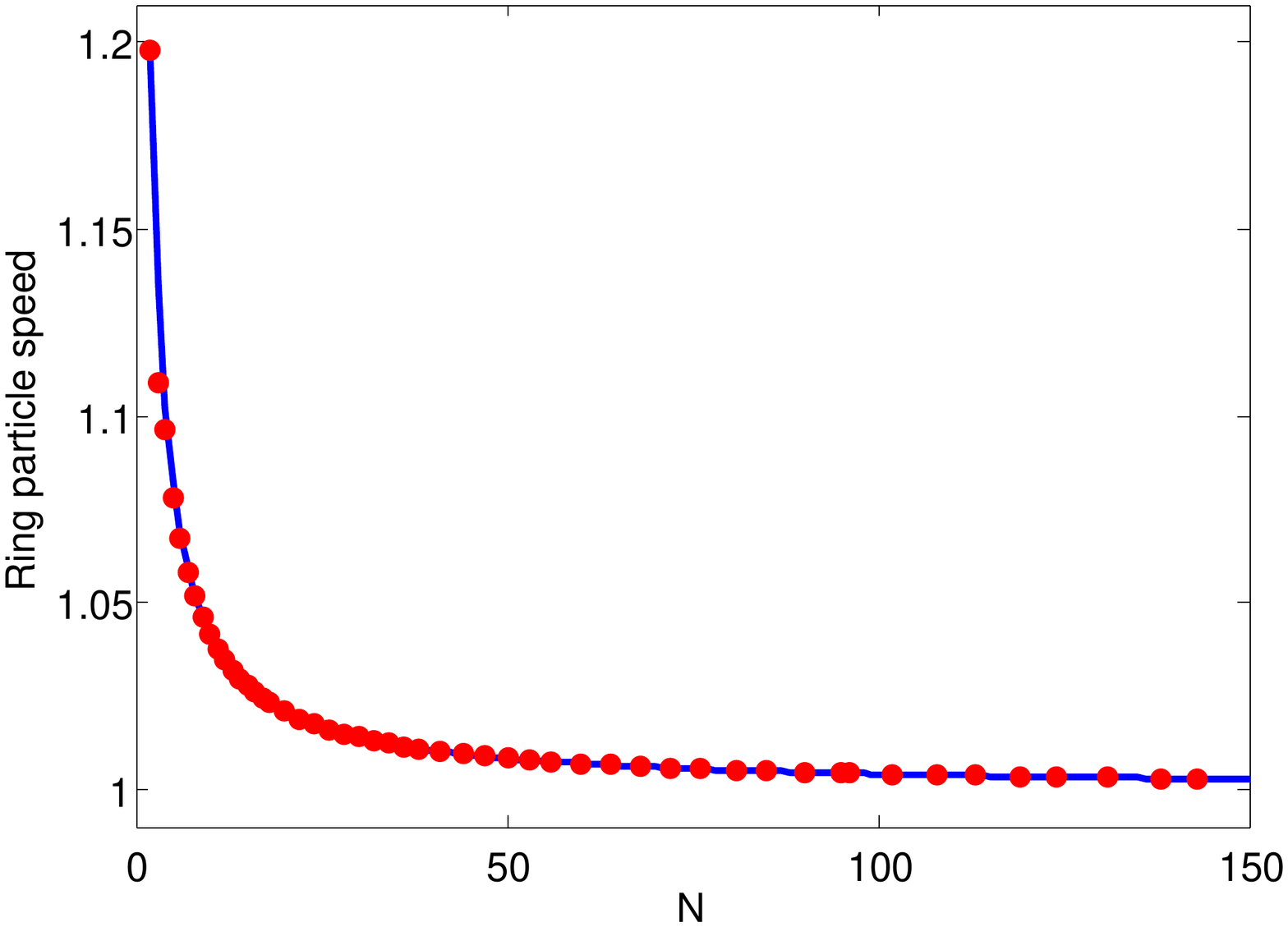}
\end{subfigure}
\caption{Radius and average speed of agents in the ring state for 
different values of swarm size $N$. Red dots indicate numerical 
simulations and the blue lines are predicted by theory. Parameter 
values: $a = 1$, $a_L = 0$, $\tau = 1$, and $\kappa_i = 1$ for all $i$.}
\label{fig:homogeneous_swarm}
\end{figure}

\begin{figure}[ht!]
 \centering
\includegraphics[width=0.5\textwidth]{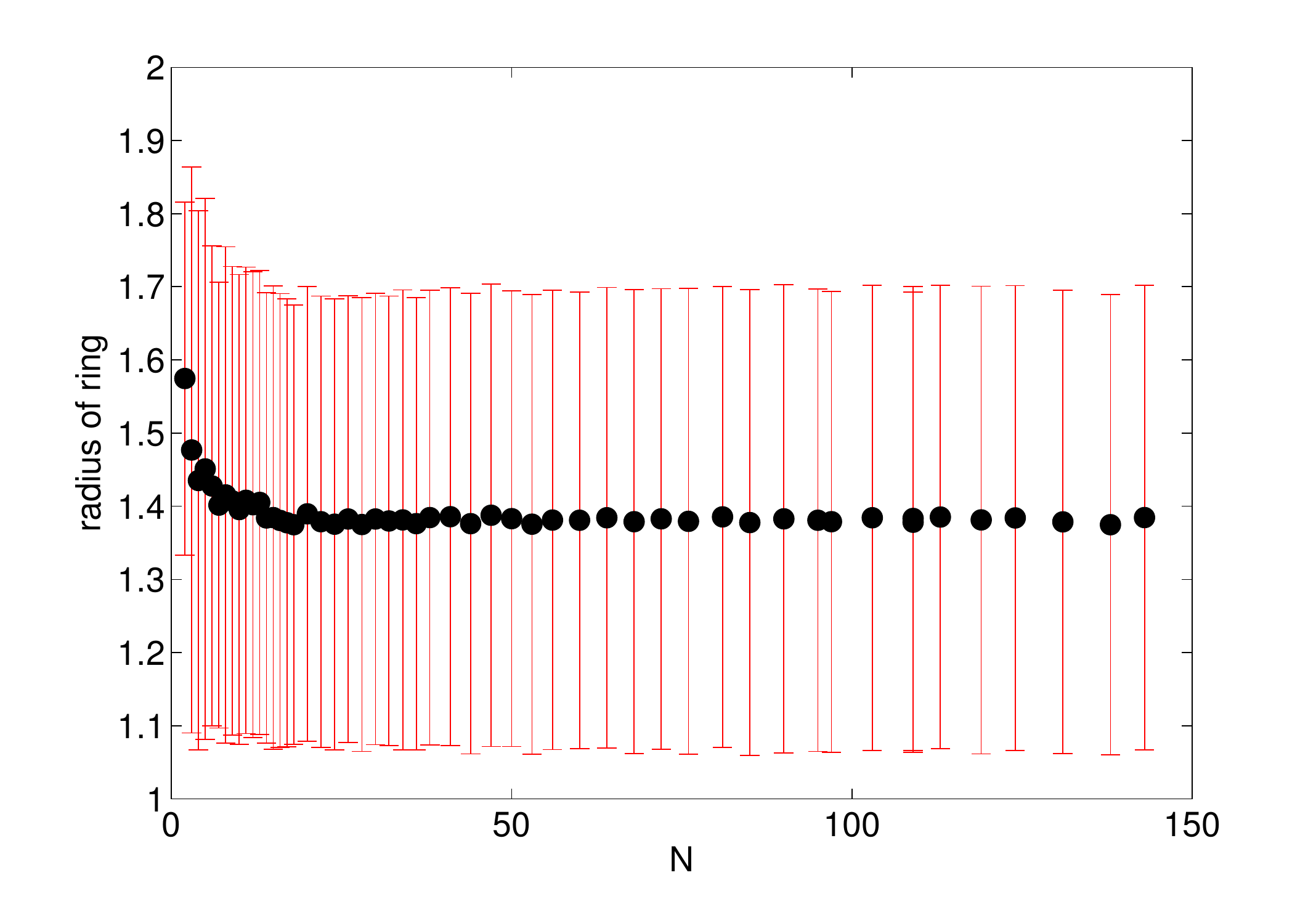}\\
\includegraphics[width=0.5\textwidth]{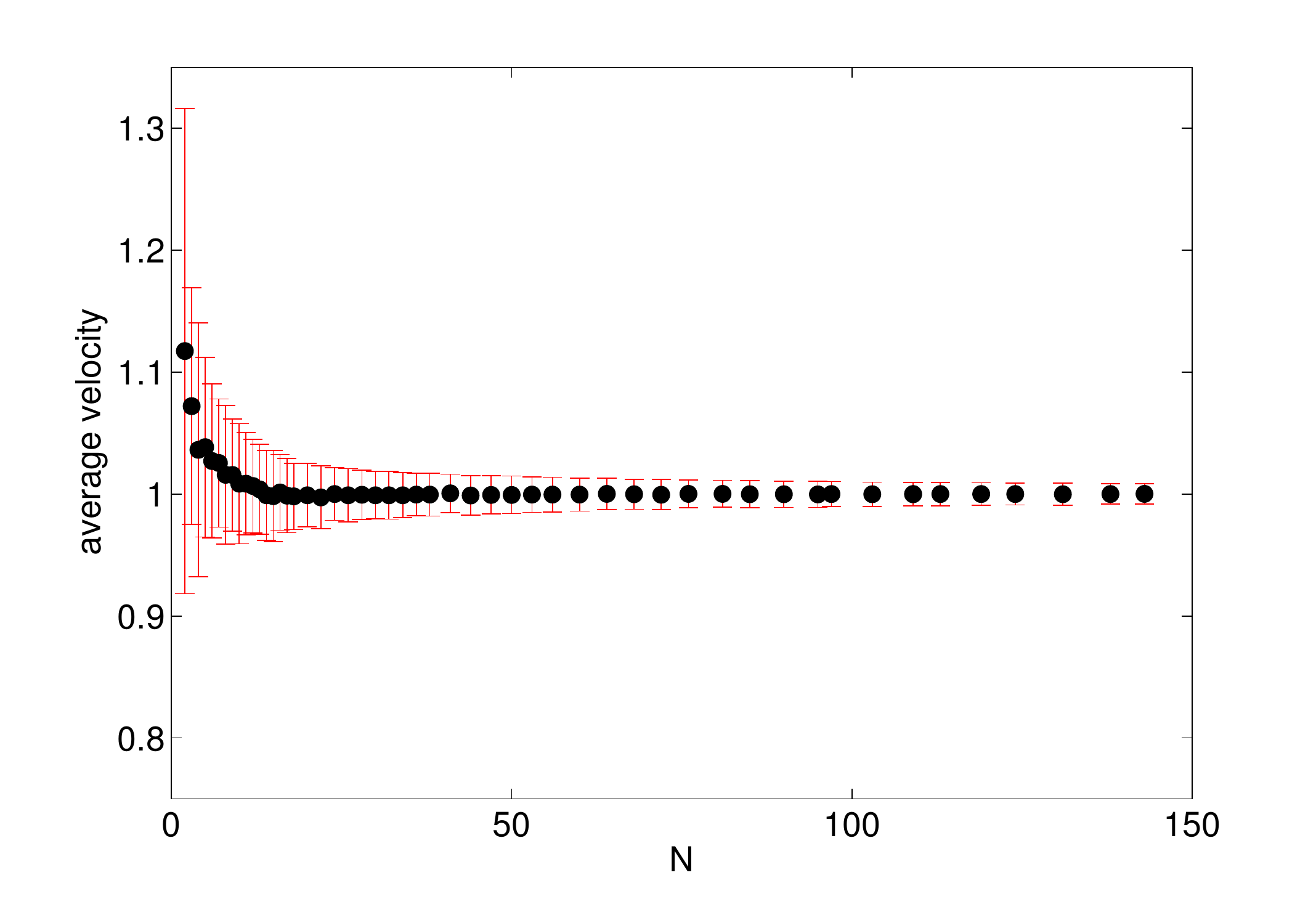}
\caption{Radius and average speed of agents in the ring state for 
different values of swarm size $N$ with uniformly distributed
values of $\kappa$. Parameter values: $a = 1$, $a_L = 
0$, $\tau = 1$, and $\kappa_i$ uniformly distributed on $[0.2,1]$.}
\label{fig:heterogeneous_swarm}
\end{figure}


\section{Conclusion}
In this paper we have used curve following as a sample application for the use of collectively moving autonomous agents. We have shown how the naturally-emerging collective motions of interacting autonomous agents can be exploited to segregate agents with different dynamical properties, even when they follow the same overall behaviors as other agents in the swarm (in our case, we separated out agents with lower acceleration factors, corresponding to depleted battery state or mechanical failure).  

Furthermore, we have analyzed collective motions of delay-coupled
heterogeneous agents. We tested the limits of the commonly-used thermodynamic
limit for modeling swarm populations by considering the effects of finite
swarm size on time to converge to a given pattern (in this case, the ring
state), and by comparing the ring state radius and circulating velocity with
theoretical predictions based on the thermodynamic model. We have verified
that the theoretical predictions of the thermodynamic limit hold very well for
large (100+ agents) and medium (20+ agents) swarms, but break down for smaller
numbers of agents. In this few-agent limit, the swarm often does not converge to the
expected ring, and we observe the emergence of more exotic collective periodic patterns. 

In the current work, we have assumed that the agents in the swarm are globally coupled. This is generally not feasible in swarms of more than a few agents, on account of communication bandwidth requirements. In future work, we will relax this assumption to consider less than fully connected swarms. We will also consider the effects of inter-agent repulsion on the collective dynamics. 

Our work represents an important link between theory of aggregate systems (generally developed in the thermodynamic limit) and practical applications of swarming systems (which generally contain few agents).

\section*{Acknowledgments}

This research was performed while KS and CRH held a National Research Council Research Associateship Award at the U.S. Naval Research Laboratory.  LMTR is a post doctoral fellow at Johns Hopkins University supported by the National Institutes of Health. This research
is funded by the Office of Naval Research contract no. N0001412WX2003 and the Naval Research 
Laboratory 6.1 program contract no. N0001412WX30002.


\end{document}